\documentstyle[twocolumn,aps,prd]{revtex}
\tighten
\begin{document}
\draft

\title{A 3+1 covariant suite of Numerical Relativity Evolution Systems}
\author{C.~Bona, T.~Ledvinka$^{\dag}$ and C.~Palenzuela}
\address{
Departament de Fisica, Universitat de les Illes Balears,
Ctra de Valldemossa km 7.5, 07071 Palma de Mallorca, Spain\\
$^{\dag}$Institute of Theoretical Physics, Faculty of Mathematics
and Physics, Charles University, V Hole\v{s}ovi\v{c}k\'ach 2, 180
00 Prague 8, Czech Republic}

\maketitle

\begin{abstract}
A suite of three evolution systems is presented in the framework
of the 3+1 formalism. The first one is of second order in space
derivatives and has the same causal structure of the
Baumgarte-Shapiro-Shibata-Nakamura (BSSN) system for a suitable
choice of parameters. The second one is the standard first order
version of the first one and has the same causal structure of the
Bona-Masso system for a given parameter choice. The third one is
obtained from the second one by reducing the space of variables
in such a way that the only modes that propagate with zero
characteristic speed are the trivial ones. This last system has
the same structure of the ones recently presented by Kidder,
Scheel and Teukolski: the correspondence between both sets of
parameters is explicitly given. The fact that the suite started
with a system in which all the dynamical variables behave as tensors
(contrary to what happens with BSSN system) allows one to keep the
same parametrization when passing from one system to the next
in the suite. The direct relationship between each parameter and a
particular characteristic speed, which is quite evident in the second
and the third systems, is a direct consequence of the manifest 3+1
covariance of the approach.
\end{abstract}

\section{Introduction}

In a recent article \cite{Cornell01} (which will be hereafter
referred to as the KST paper), a wide class of hyperbolic
first order formalisms has been studied with a view to Numerical
Relativity applications. All these formalisms do use the well
known $3+1$ decomposition of spacetime
\begin{eqnarray}\label{metric4D}
    ds^2 &=& - \alpha^2\;dt^2  \nonumber \\
    &+& \gamma_{ij}\;(dx^i+\beta^i\;dt)\;(dx^j+\beta^j\;dt)
     \;\;\;\;i,j=1,2,3
\end{eqnarray}
which allows one to use Einstein's field equations as an evolution
system for the metric (\ref{metric4D}), namely
\begin{eqnarray}
   (\partial_t -{\cal L}_{\beta}) \gamma_{ij} &=& -2\;\alpha\;K_{ij}
\label{evolve_metric} \\
   (\partial_t -{\cal L}_{\beta}) K_{ij} &=& -\nabla_i\alpha_j
\label{evolve_K} \\
    &+& \alpha\; [{}^{(3)}R_{ij}-2K^2_{ij}+trK\;K_{ij}-{\hat{R}}_{ij}]
\nonumber
\end{eqnarray}
(ADM system). The KST paper describes many ways of obtaining a
first order version of the second order ADM system by introducing
a number of arbitrary parameters. Different parameter choices
correspond to different ways of using the energy and momentum
constraints
\begin{eqnarray}
   ^{(3)}R - tr(K^2) + (trK)^2 &=& 2\;\tau
\label{energy_constraint}  \\
   \nabla_k\:{K^k}_{i}-\partial_i(trK) &=& S_i
\label{momentum_constraint}
\end{eqnarray}
to modify the structure of the resulting systems. The goal is to
get well posed systems, so that existence, unicity and stability
of the solutions can be ensured \cite{F82}. In this sense, the KST
paper can be understood as a generalization of previous works by other
groups \cite{FR94,FR96,AY99}, which have in common the fact that
the only independent quantities are the metric coefficients and
their first derivatives.

A similar approach can be used to generalise the hyperbolic first
order formalisms proposed by the Palma group \cite{BM92,BM95}, as
suggested in Ref. \cite{Bona99}. The main difference with the
formalisms studied in the KST paper is that in this case there
are three supplementary variables, independent of the metric
derivatives, whose evolution equations are obtained by reversing
the order of space and time derivatives in the momentum
constraint (\ref{momentum_constraint}). This use of the momentum
constraint to evolve three supplementary variables can also be
found in another context \cite{SN95,BS99}, where a second order
system of the form (\ref{evolve_K}) is considered (BSSN system).

To summarize, we have different formalisms in which the
constraints (\ref{energy_constraint}, \ref{momentum_constraint})
are used to modify the structure of the ADM evolution system.
The momentum constraint is used in some cases to evolve
supplementary quantities, either keeping the system to be of
second order \cite{SN95,BS99} or getting first order versions
\cite{BM92,BM95}. But the same momentum constraint is used instead
in other cases to modify the first order versions without introducing
extra quantities \cite{FR94,FR96,AY99}.

The purpose of the present article is to point out the strong relationship
among all these formalisms from the theoretical point of view. With this
aim, we propose a new covariant framework in which the momentum constraint
is used to evolve a vector quantity, so that the properties of the resulting
systems are independent of the choice of space coordinates.

The paper is organised as follows: In Section II, we introduce the
vector quantity $Z_i$, which will replace the corresponding
non-vector supplementary quantities in \cite{BM92,BM95,Bona99,SN95,BS99}.
The original ADM system is then generalised by making use of the vector
$Z_i$ and the energy constraint (\ref{energy_constraint}). In Section III,
the resulting second order system (which will be hereafter referred to as
system A) is compared with the BSSN system \cite{SN95,BS99}.

In section IV, we obtain a first order version of system A (which will be
hereafter referred to as system B), which is compared with the formalisms
presented in \cite{BM92,BM95}. Finally, in section V, system B
is further modified in order to be compared with the formalisms discussed in
the KST paper.

\section{Integrating the momentum constraint}
Let us define the vector $Z^i$ as follows:
\begin{equation}\label{evolve_Zs}
   (\partial_t -{\cal L}_{\beta})Z_i =
    \alpha\;[\nabla_j\:({K_i}^{j}-{\delta_i}^j\;trK) - S_i]
\end{equation}
so that it is clear that $Z_i$ vanishes for physical solutions, which can
be defined as the ones verifying both the energy and momentum constraints
(\ref{energy_constraint}, \ref{momentum_constraint}). The
"zero" vector $Z_i$ can then be considered as a good covariant
measure of the deviation of any solution $(\alpha,\gamma_{ij},K_{ij})$ of
the ADM system from the submanifold of physical solutions, as far as
the momentum constraint is concerned. This means that the cumulative
effect of the differential constraint (\ref{momentum_constraint}) can be
allowed for by looking at the simple algebraic equation
\begin{equation}\label{Zzero}
  Z_i=0
\end{equation}
which is to be regarded as the integral version of (\ref{momentum_constraint}).

But these $Z_i$, besides being useful auxiliary quantities, can be
also considered as supplementary dynamical variables. This is the
kind of approach proposed in Ref. \cite{FR99}, where a
"$\lambda$-system" was proposed containing as much as 40
supplementary quantities to be evolved by using the constraints. We
will limit ourselves here to only three quantities $Z_i$ (corresponding
to $\lambda_i$ in \cite{FR99}), the ones related with the momentum
constraint. We will use these quantities $Z_i$, together with the energy
constraint (\ref{energy_constraint}) to implement
covariant quasilinear modifications of the principal part of the ADM
system, without affecting the submanifold of
physical solutions (the ones verifying (\ref{energy_constraint}, \ref{Zzero})).
We will call "system A" the most general second order evolution system
obtained in this way, namely
\begin{eqnarray}\label{evolve_K_par}
   (\partial_t -{\cal L}_{\beta}) K_{ij} &=& -\nabla_i\alpha_j \\
   &+& \alpha\; [{}^{(3)}R_{ij}-2K^2_{ij}+ trK\;K_{ij}-{\hat{R}}_{ij}]
\nonumber  \\
    &-& \frac{\alpha\;n}{4}[R-tr(K^2)+tr^2(K)-2\;\tau]\;\gamma_{ij}\nonumber \\
    &+& \frac{\alpha\;\mu}{2} (\nabla_i Z_j + \nabla_j Z_i)
     - \frac{\alpha\;\nu}{2} (\nabla \cdot Z)\;\gamma_{ij} \nonumber
\end{eqnarray}
where $\mu$, $\nu$ and $n$ are arbitrary parameters.

The modified evolution system (\ref{evolve_Zs},\ref{evolve_K_par})
(system A) is of a mixed type: first order in ($Z_i,
K_{ij}$) but second order in ($\alpha, \gamma_{ij}$). In this
sense, it is very similar to the BSSN system \cite{SN95,BS99}.
In order to compare these systems, let us state some definitions:\\

a) Two evolution systems will be said to be equivalent if one can
be obtained from the other by a change of variables. It is clear that
two equivalent systems have the same space of solutions.

b) Two evolution systems will be said to be quasiequivalent when their
principal parts are equivalent. It follows that two quasiequivalent
systems have the same causal structure.\\

For instance, if we introduce "damping terms" to modify eq.
(\ref{evolve_Zs}) (as proposed in \cite{FR99} with a view to
Numerical Relativity applications), we are not affecting the principal
part (damping terms are algebraic in $Z_i$), so that the resulting
"damped" system  would be quasiequivalent to system A. Also, if we introduce
additional terms of the form $Z_iZ_j$ or $Z^2\gamma_{ij}$ in
(\ref{evolve_K_par}), the resulting modified system would be
quasiequivalent to system A as well.

On the contrary, different values of the parameters $\mu$, $\nu$, $n$
in (\ref{evolve_K_par}) do correspond to different forms of the principal
part, so that they would lead to different evolution systems which are
not quasiequivalent, even if they have (as they do) the same submanifold
of physical solutions. In this context, it is natural to use system A
as a framework to compare different formalisms that have been proposed
with a view to Numerical Relativity applications, where the causal
structure of the evolution system is important to ensure the well posedness
of the problem, namely the existence, unicity and stability of the
solutions.

We can also take advantage of the covariance of system A under
coordinate transformations of the generic form:
\begin{eqnarray}
    t'   &=& f(t) \\
    {x^{i}}' &=& g(x^i,t) \nonumber
\end{eqnarray}
(3+1 covariance) so that we can choose $\beta^i=0$ (normal coordinates)
in what follows without any loss of generality.

\section{Comparing with the BSSN system}

Let us consider here the system introduced in by Shibata and Nakamura
\cite{SN95} and Baumgarte and Shapiro \cite{BS99} (BSSN system).
The metric coefficients $\gamma_{ij}$ are expressed in terms of a
conformal metric:
\begin{equation}\label{conformal_metric}
  {\tilde{\gamma}}_{ij} = e^{-4\;\phi}\;\gamma_{ij}
\end{equation}
with unit determinant, so that
\begin{equation}\label{conformal_factor}
   e^{4\;\phi} ={\gamma}^{1/3}=[det(\gamma_{ij})]^{1/3}
\end{equation}
The second fundamental form $K_{ij}$ is then decomposed into its
trace and trace-free components, namely
\begin{eqnarray}
  K &=& \gamma^{ij}\;K_{ij}
\label{traceK} \\
  {\tilde{A}}_{ij} &=& e^{-4\;\phi}\;(K_{ij}-
  \frac{1}{3}\; K\;\gamma_{ij})
\label{tracelessK}
\end{eqnarray}
so that the principal part of their evolution equations can be written
as \cite{BS99}:
\begin{eqnarray}
   \partial_t  K &=& - \gamma^{ij}\;\nabla_{i}\; \alpha_j + ...
\label{evolve_traceK} \\
   \partial_t {\tilde{A}}_{ij} &=& e^{-4\; \phi}\;
   {(-\nabla_{i}\;\alpha_j + \alpha\; R_{ij})}^{TF}+...
\label{evolve_tracelessK}
\end{eqnarray}
where the superscript TF denotes the trace-free part of a tensor.

The principal part of the Ricci tensor in (\ref{evolve_tracelessK}) is
decomposed as follows \cite{BS99}:
\begin{eqnarray}\label{conformal_Ricci}
   R_{ij} &=& -2\; {{\partial}^2}_{ij} \phi -
    \frac{1}{2}\;{\tilde{\gamma}}^{rs}\;{\tilde{\gamma}}_{ij,rs} +
   {\tilde{\gamma}}_{k(i}\; \partial_{j)}\;{\tilde{\Gamma}}^k + ...
\end{eqnarray}
where the ``conformal connection'' quantities ${\tilde{\Gamma}}^i$
can be defined as:
\begin{equation}\label{Gs}
   {\tilde{\Gamma}}^i = -{{\tilde{\gamma}}^{ij}}_{\;,j}
\end{equation}
Up to now we have only performed a transformation of variables of
the original ADM system. But notice that the system
(\ref{evolve_traceK}, \ref{evolve_tracelessK}) contains seven
evolution equations, compared with the six equations in
(\ref{evolve_K}). The extra quantity is the trace of
${\tilde{A}}_{ij}$, which is supposed to vanish by construction.
This opens the way to two different approaches:

a) to consider $tr(\tilde{A})$ as an independent variable, which
can take non-zero values during numerical evolution. The equation
\begin{equation}
    tr(\tilde{A})=0 \nonumber
\end{equation}
is to be considered as an additional constraint which can be used
to monitor the accuracy of the simulation (free evolution
approach).

b) to enforce the vanishing of $tr(\tilde{A})$, by removing
systematically any contribution that could arise from truncation
errors in the discretization of equation
(\ref{evolve_tracelessK}), so that $tr(\tilde{A})$ is not
considered as a new dynamical variable (constrained evolution
approach)

In what follows, we will adopt the second approach, so that the
system (\ref{evolve_traceK}, \ref{evolve_tracelessK}) is then
quasiequivalent to the ADM system.

The BSSN system does consider the ${\tilde{\Gamma}}^i$ as
independent dynamical variables, so that the full set of
independent BSSN variables is given by
\begin{equation}\label{BSSN_vars}
   u=\{ \alpha,\; \phi,\; {\tilde{\gamma}}_{ij},\;K,\;
      {\tilde{A}}_{ij},\; {\tilde{\Gamma}}^i \}\;\;,
\end{equation}
where we must remember that both the trace of ${\tilde{A}}_{ij}$
and the determinant of ${\tilde{\gamma}}_{ij}$ do have fixed,
non-dynamical, values. An evolution equation for the
${\tilde{\Gamma}}^i$ could be derived \cite{BS99} by permuting a
time derivative with the space derivative in (\ref{Gs})
\begin{equation}\label{evolve_Gs_}
   \partial_t {\tilde{\Gamma}}^i = -2\; \alpha\;
   {{\tilde{A}}^{ij}}_{,j} + ...
\end{equation}
But the BSSN system uses instead the momentum constraint
(\ref{momentum_constraint}) to transform eq. (\ref{evolve_Gs_})
into
\begin{equation}\label{evolve_Gs}
   \partial_t {\tilde{\Gamma}}^i = -\frac{4}{3}\;
            \alpha\; {\tilde{\gamma}}^{ij}\; K_{,j}+...
\end{equation}

Now we are in position to compare the BSSN system with the one introduced
in the previous section (system A), with a set of variables given by
\begin{equation}\label{A_vars}
   u=\{ \alpha,\; \gamma_{ij},\;K_{ij},\; Z_i \}
\end{equation}
The conformal decomposition (\ref{conformal_metric},
\ref{conformal_factor}, \ref{traceK}, \ref{tracelessK}) can be
easily performed on system A as well. The key point is then the
passage from (\ref{evolve_Gs_}) to the BSSN evolution equation
(\ref{evolve_Gs}). In the framework of system A, the momentum
constraint is just the right-hand-side of the evolution equation
for $Z_i$. It follows that the use of the momentum constraint to
transform (the principal part of) the right-hand-side of
(\ref{evolve_Gs_}) into (\ref{evolve_Gs}) does correspond to the
following transformation of ${\tilde{\Gamma}}^i$:
\begin{equation}\label{Gs2}
  {\tilde{\Gamma}}_i = -{\tilde{\gamma}}_{ik}\;{{\tilde{\gamma}}^{kj}}_{\;,j}
                       + 2\; Z_i
\end{equation}
which is obviously consistent with the previous expression
(\ref{Gs}) if we remember that $Z_i=0$ for the physical solutions.

Now we can compare easily (\ref{evolve_tracelessK}) with the
trace-free part of (\ref{evolve_K_par}): it follows that they coincide
if and only if $\mu=2$. The same comparison can be performed again
between (\ref{evolve_traceK}) and the trace part of (\ref{evolve_K_par}):
it follows that they coincide if and only if $\nu=n=\frac{4}{3}$.

To summarize, we have shown then that the BSSN system is quasiequivalent
to system A for the parameter choice
\begin{equation}\label{BSSN_pars}
       \mu=2,\;\; \nu=n=4/3
\end{equation}
provided that the ``conformal connection'' quantities ${\tilde{\Gamma}}^i$
in the BSSN system are related with the vector $Z_i$ by equation
(\ref{Gs2}).

\section{A first order evolution system}

A first order version of the system A can be obtained in the standard
way by considering the first space derivatives
\begin{eqnarray}
    A_k &=& \partial_k (\ln \alpha)
\label{As} \\
    D_{kij} &=& \frac{1}{2}\; \partial_k \gamma_{ij}
\label{Ds}
\end{eqnarray}
as independent dynamical quantities whose evolution equations are given
by
\begin{eqnarray}
     \partial_t A_k + \partial_k(\alpha\;Q) &=& 0
\label{evolve_As} \\
   \partial_t D_{kij} + \partial_k(\alpha\; K_{ij}) &=& 0
\label{evolve_Ds}
\end{eqnarray}
where the quantity $Q$, defined as
\begin{equation}\label{evolve_lapse}
     \partial_t \ln \alpha = - \alpha\;Q
\end{equation}
can be related with $trK$ in order to fix the time coordinate gauge,
namely
\begin{equation}\label{Q}
    Q=f\;trK
\end{equation}
where $f$ is an arbitrary function of the lapse $\alpha$.

This allows to transform system A into a first order evolution system
for the set of variables
\begin{equation}\label{B_vars}
   u=\{ \alpha,\; \gamma_{ij},\;K_{ij},\; A_i,\; D_{kij},\; Z_i \}
\end{equation}
with a non-trivial principal part given by (\ref{evolve_Zs},
\ref{evolve_As}, \ref{evolve_Ds}) and
\begin{equation}\label{evolve_K_principal}
   \partial_t K_{ij} + \partial_k (\alpha\;{\lambda^k}_{ij}) = ...
\end{equation}
where we have noted
\begin{eqnarray}\label{flux_K}
   {\lambda^k}_{ij} = {D^k}_{ij}
   &+& \frac{1}{2}{\delta^k}_i(A_j + D_j - 2\;{D_{rj}}^r - \mu\;Z_j)
\\
   &+& \frac{1}{2}{\delta^k}_j(A_i + D_i - 2\;{D_{ri}}^r - \mu\;Z_i)
\nonumber \\
  &-& \frac{n}{2}(D^k-{D^{rk}}_r)\gamma_{ij}
      + \frac{\nu}{2}Z^k\;\gamma_{ij} \nonumber
\end{eqnarray}
and
\begin{equation}
    D_k =\gamma^{ij}\;D_{kij} \:\:.
\end{equation}
This first order system will be called system B in what follows.

System B is very similar to the Bona-Masso system \cite{BM92,BM95}
(we will follow here the notation of Ref. \cite{BM95} to avoid confusion),
which is expressed in terms of the set of variables
\begin{equation}\label{BM_vars}
   u=\{ \alpha,\; \gamma_{ij},\;K_{ij},\; A_i,\; D_{kij},\; V_i \}
\end{equation}
where the quantities $V_i$ can be defined as
\begin{equation}\label{Vs}
    V_i = D_{i} - {D_{ji}}^j
\end{equation}
An evolution equation for the $V_i$ could be derived by combining
(\ref{Vs}) with the evolution equations (\ref{evolve_Ds}) for the
$D_{kij}$
\begin{equation}\label{evolve_Vs_}
    \partial_t V_i = \alpha\; \partial_j \;({K_i}^j-trK\; {\delta_i}^j)
    + ...
\end{equation}
but the Bona-Masso system uses instead the momentum constraint to
transform (\ref{evolve_Vs_}) into an evolution equation with a
trivial principal part, namely
\begin{equation}\label{evolve_Vs}
    \partial_t V_i = ...
\end{equation}
The remaining evolution equations in the Bona-Masso system are identical
to those in system B, with the replacement of (\ref{flux_K}) by
\begin{eqnarray}\label{flux_K2}
   {\lambda^k}_{ij} &=& {D^k}_{ij}
   + \frac{1}{2}{\delta^k}_i(A_j + 2\;V_j - D_{j}) \\
   &+& \frac{1}{2}{\delta^k}_j(A_i + 2\;V_i - D_{i})
    - \frac{n}{2}\; V^k\; \gamma_{ij} \nonumber
\end{eqnarray}
In order to compare system B with the Bona-Masso system, the key
point is again the passage from (\ref{evolve_Vs_}) to the
Bona-Masso evolution equation (\ref{evolve_Vs}). The use of the
momentum constraint to transform (the principal part of) the
right-hand-side of (\ref{evolve_Vs_}) into (\ref{evolve_Vs})) does
correspond to the following transformation of $V_i$:
\begin{equation}\label{Vs2}
    V_i = D_{i} - {D_{ji}}^j - Z_i
\end{equation}
which is again consistent with (\ref{Vs}) because $Z_i$ vanishes for physical
solutions.

Now we can substitute (\ref{Vs2}) into (\ref{flux_K2}) and compare  with
(\ref{flux_K}): it follows that they coincide if and only if
\begin{equation}\label{BM_pars}
       \mu=2,\;\; \nu=n
\end{equation}
To summarize, we have shown that the Bona-Masso system is quasiequivalent
to system B for the parameter choice (\ref{BM_pars}), provided that the
supplementary quantities $V_i$ in the Bona-Masso system are related with
the vector $Z_i$ by equation (\ref{Vs2}).
As a consequence of this fact, system B inherits (for the parameter choice
(\ref{BM_pars})) the causal structure of the Bona-Masso system, which is
known to be strongly hyperbolic (real characteristic speeds and a complete
set of eigenfields \cite{F82}) if and only if $f>0$.

\section{Reducing variables space}

System B has two sets of non-trivial standing eigenmodes (the ones
with zero characteristic speed in normal coordinates). The first
one is related with gauge evolution, namely
\begin{equation}\label{zeromode1}
     A_i -f\; D_i
\end{equation}
while the second one is the combination (\ref{Vs2}), related with
the momentum constraint. We will see how we can use this fact to
reduce the set of independent dynamical variables (\ref{B_vars})
of the system B to the smaller one used in the KST paper, namely
\begin{equation}\label{KST_vars}
   u=\{ \gamma_{ij},\;K_{ij},\; d_{kij} \}
\end{equation}
where, in spite of the fact that
\begin{equation}\label{d_D}
    d_{kij}= 2 \; D_{kij}
\end{equation}
we will keep using here the KST original notation to avoid
confusion.

First of all, one can explicitly integrate the relationship
(\ref{evolve_lapse}, \ref{Q}) between the lapse and the volume
element in the case $f=2\; \sigma=constant$, namely
\begin{equation}\label{densitized_lapse}
    \partial_t (\alpha\; \gamma^{-\sigma}) = 0
\end{equation}
so that the value of $\alpha$ can be defined in terms of $\gamma$
for every initial condition (densitized lapse). The same thing can
be done with $A_i$ and $D_i$, so that
\begin{equation}\label{densitized_AsDs}
     A_i \equiv 2\; \sigma\; D_i + ...
\end{equation}
The list of dynamical variables is then reduced from
(\ref{B_vars}) to
\begin{equation}\label{EC_vars}
   u=\{ \gamma_{ij},\;K_{ij},\; D_{kij},\; Z_i \}
\end{equation}
where the gauge-related standing modes (\ref{zeromode1}) have just
disappeared from the dynamical system.

In doing this, we have restricted ourselves to the $f=constant$
case, in spite of the fact that this case is known \cite{AM98} to
be prone to numerical gauge instabilities if $f \neq 1$. We do
that with a view to mimic the KST system, where $\sigma$ is
defined to be a constant and, as a consequence of it, only
examples with $\sigma=1/2$ ($f=1$) are provided. In constrast,
system B would allow more general gauge choices, like the "1+log"
one ($f=2/\alpha$) which is frequently used in Numerical
Relativity.

Disposing of the supplementary variables $Z_i$ in (\ref{EC_vars})
is not so easy. We will take as a guide the KST evolution
equations for $\gamma_{ij}$ and $K_{ij}$, which are very similar
to the corresponding ones in system B: we need just to replace
(\ref{flux_K}) by
\begin{eqnarray}\label{flux_K_KST}
   2\; {\lambda^k}_{ij} &=& {d^k}_{ij}
   + \frac{1}{2}{\delta^k}_i[(1+2\; \sigma)\; d_j - 2\;{d_{rj}}^r] \\
   &+& \frac{1}{2}{\delta^k}_j[(1+2\; \sigma)\; d_i - 2\;{d_{ri}}^r]
  - \frac{n}{2}(d^k-{d^{rk}}_r)\gamma_{ij} \nonumber
\end{eqnarray}
where we have taken $\zeta=-1,\; \gamma=-n/2$ in the original KST
expression to reproduce (\ref{flux_K}) more closely.

The main difference between system B and the KST one comes instead from
the evolution equations for $d_{kij}$, namely
\begin{eqnarray}\label{evolve_ds}
   &\partial_t& d_{kij} = - 2\; \alpha\; \partial_k K_{ij}
   + \chi\; \alpha\; \gamma_{ij}(\partial_r {K^r}_k -\partial_k trK) \\
 &+& \frac{\eta}{2}\; \alpha [\gamma_{ki} (\partial_r {K^r}_j - \partial_j trK)
 + \gamma_{kj} (\partial_r {K^r}_i - \partial_i trK)] + ... \nonumber
\end{eqnarray}
where $\eta, \chi$ are arbitrary parameters. These equations are
obtained by using the momentum constraint
(\ref{momentum_constraint}) to modify the standard equations
(\ref{evolve_Ds}) which were used in system B. The key point again
is to realize that the use of the momentum constraint to transform
(the principal part of) the right-hand-side of (\ref{evolve_Ds})
into that of (\ref{evolve_ds}) amounts, allowing for
(\ref{evolve_Zs}), to modify the relationship (\ref{d_D})
between $D_{kij}$ and $d_{kij}$ in the following way:
\begin{equation}\label{d_D2}
       2\; D_{kij} = d_{kij} - \eta\; \gamma_{k(i}\; Z_{j)}
       - \chi\; Z_k \gamma_{ij}
\end{equation}
which is consistent again with (\ref{Ds}) because $Z_i$ vanishes
for physical solutions. A straightforward substitution of
(\ref{densitized_AsDs}) and (\ref{d_D2}) into (\ref{flux_K}) shows
that the vector $Z_i$ disappears from the principal part if we
just relate the free parameters in system B with those of the KST
system as follows:
\begin{eqnarray}\label{pars_B_KST}
   f &=& 2\; \sigma \\
  \nu &=& \chi - n(\chi - \eta/2)
\nonumber \\
  \mu &=& \eta - \chi/2 - \sigma (\eta + 3\; \chi)
\nonumber
\end{eqnarray}
so that the Einstein-Christoffel system \cite{AY99} (corresponding to
$\sigma=1/2$, $\eta = 4$, $\gamma=\chi=0$ in the KST paper) is recovered
when
\begin{equation}\label{EC_pars}
       f=1, \;\; \mu=2,\;\; \nu=n=0
\end{equation}

To summarize, we have shown here that the KST systems are quasiequivalent
(in the case $\zeta=-1$) to a first order system which can be obtained
by reducing dynamical variables in system B to the minimal set
(\ref{KST_vars}). The reduced system will be called system B' in what
follows. The four parameters of system B' can be expressed in terms
of those of the KST system by (\ref{pars_B_KST}), where we must
remember that $n=-2\; \gamma$.

\section{Concluding remarks}

The evolution systems A, B and B' presented in this paper
are closely related in a transparent way, up to the point that we have
been able to keep the same set of parameters ($f, \mu, \nu, n$) in
the three cases. The fact that, for suitable parameter choices, they
are quasiequivalent respectively to the BSSN \cite{SN95,BS99},
Bona-Masso \cite{BM92,BM95} and Einstein-Christoffel \cite{AY99}
systems points out the strong relationship between these systems,
proposed independently in different contexts by different authors,
all of them extremely useful in Numerical Relativity applications.

There are at least two specific points in which the suite A, B, B'
can improve our understanding of the previous systems. The first one
is that the supplementary quantities $Z_i$ in systems A and B are
three-dimensional vectors, contrary to what happens with their
counterparts ${\tilde{\Gamma}}^i, V_i$ respectively, which do not
have tensor behaviour. Then, systems A and B can be
used to build up covariant counterparts of the second order BSSN
\cite{SN95,BS99} and first order Bona-Masso \cite{BM92,BM95} systems,
respectively.

The second point can be seen when analysing the list of non-trivial
characteristic speeds in the KST systems, namely
\begin{equation}\label{KST_speeds}
   \{ \pm 1,\; \pm c_1,\; \pm c_2,\; \pm c_3 \}
\end{equation}
where in our case ($\zeta=-1$) we have \cite{Cornell01}
\begin{eqnarray}\label{KST_speeds2}
   {c_1}^2 &=& 2\; \sigma \\
   {c_2}^2 &=& \frac{1}{4} (2\; \eta - 2\; \eta\; \sigma -
                \chi - 6\; \chi\; \sigma)
\nonumber \\
  {c_3}^2 &=& 1 + 2\;\gamma-\gamma\;\eta + \chi+2\;\gamma\;\chi \nonumber
\end{eqnarray}
which can be translated, allowing for (\ref{pars_B_KST}) into
\begin{equation}\label{KST_B_speeds}
     {c_1}^2 = f, \;\;  {c_2}^2=\mu/2,\;\;  {c_3}^2=1+\nu-n
\end{equation}
where we have used $n=-2\; \gamma$.

Expression (\ref{KST_B_speeds}) allows to decouple the parameter set
in a way such that every parameter is related with just one
characteristic speed, making the physical interpretation more
transparent: the gauge parameter f is associated only with $c_1$
(gauge speed), the parameter $\mu$ (which appears in the terms affecting
to the longitudinal part of $K_{ij}$) is associated only with $c_2$,
whereas the two parameters $\nu$ and $n$ appearing in the transverse
trace terms are related with $c_3$ (notice that there is some degree
of redundancy between $\nu$ an $n$).

It is not surprising then that in all the particular sets of parameters
corresponding to previous works \cite{FR96,AY99,BM92,BM95,Bona99,SN95,BS99}
(\ref{BSSN_pars}, \ref{BM_pars}, \ref{EC_pars}) one has
\begin{equation}\label{equal_pars}
     \mu=2,\;\; \nu=n
\end{equation}
which amounts to the ``physical speed'' requirement for the degrees
of freedom not related with the gauge (${c_2}^2={c_3}^2=1$). We
are aware that in the case of the BSSN system we can not speak
properly about characteristic speeds, but the close relationship
between systems A and B can bring some light even in this case
(see for instance Ref. \cite{AB99} for a similar approach).

The question arises of wether systems A, B, B' will perform any
better than their well known counterparts (BSSN, Bona-Masso, KST)
in numerical simulations. We have done preliminary tests,
comparing system B against the Bona-Masso one in black-hole
spacetimes. Our results indicate that the Bona-Masso code performs
better, because of the fact that the additional quantities $V_i$
evolve with an ordinary differential equation (\ref{evolve_Vs}),
avoiding the truncation errors inherent to the discretization of
the partial derivatives that appear instead in the evolution
equation (\ref{evolve_Zs}) for $Z_i$. The fact that $Z_i$ is a true
tridimensional vector does not help very much in that context.
Something similar happens when comparing system A with BSSN: the
evolution equation (\ref{evolve_Gs}) for the ${\tilde{\Gamma}}^i$
is still simpler than (\ref{evolve_Zs}). In our opinion, the BSSN,
Bona-Masso and KST systems are the state of the art (each one in
its class) and will be difficult to beat by any other
quasiequivalent system. One would need to provide something more
different, modifying even the causal structure of the system
(which is given by the principal part), in order to improve in a
significant way the performance (stability and accuracy) of the
current numerical simulations.\\ \\

{\em Acknowledgements: This work has been supported by the EU Programme
'Improving the Human Research Potential and the Socio-Economic
Knowledge Base' (Research Training Network Contract (HPRN-CT-2000-00137),
by the Spanish Ministerio de Ciencia y Tecnologia through the research
grant number BFM2001-0988 and by a grant from the Conselleria d'Innovacio
i Energia of the Govern de les Illes Balears.}

\bibliographystyle{prsty}

\end{document}